\documentstyle[preprint,aps]{revtex}

\begin{document}

\draft


\title{Classical and Quantum Wormholes \\
in Einstein-Yang-Mills Theory }

\author{Hongsu Kim}

\address{Department of Physics\\
Ewha Women's University, Seoul 120-750, KOREA}

\date{May, 1997}

\maketitle

\begin{abstract}
Wormhole spacetimes may be responsible for the possible loss
of quantum coherence and the introduction of additional
fundamental quantum indeterminancy of the values of constants
of nature. As a system which is known to admit such classical 
wormhole solutions, Einstein-Yang-Mills (EYM) theory
is revisited. Since the classical wormhole instanton solution
in this theory has been studied extensively thus far, in the
present work, ``quantum wormholes'' are explored. 
Namely in the context of a minisuperspace 
quantum cosmology model based on this EYM theory, ``quantum
wormhole'', defined as a state represented by a solution to the
Wheeler-DeWitt equation satisfying an appropriate wormhole
boundary condition, is discussed. Finally, it is proposed
that the minisuperspace model based on this theory in the presence
of the cosmological constant may serve as a simple yet interesting
system displaying an overall picture of entire universe's history
from the deep quantum domain all the way to the classical domain.

\end{abstract}




\centerline {\rm\bf I. Introduction}

If one identifies the spacetime metric as the relevant gravitational
field subject to the quantization, the topology of spacetime is
expected to fluctuate on scales of the order of the Planck length
$l_{p} = M^{-1}_{p}$. And of all types of conceivable spacetime
fluctuations, major concern has been the  ``wormhole configuration''
which is an object that can be loosely defined as the instanton which
is a saddle point of the Euclidean action making dominant contribution
to the topology changing transition amplitude. Then one of the most
crucial effects the wormhole (or more generally, these spacetime
fluctuations) may have on low-energy physics could be the possible
effective loss of quantum coherence [1-3]. For example, one may 
speculate the
situation where ``baby universes'' are pinched off and carry away
information. Then this kind of stereotypical information loss can
lead to an effective loss of quantum coherence as viewed by the
macroscopic observer who cannot measure the quantum state of the baby
universes. This issue of the possible loss of quantum coherence in
(mostly the semiclassical approximations of) quantum gravity stood
at the center of debates some time ago [1-3].
In a more careful and concrete analysis, however, these arguments
concerning the possible quantum incoherence
should be taken with some caution. For example, there is an argument
by Coleman [3] that in the context of ``many universe'' interpretation
(i.e., the third quantization formalism), the quantum incoherence
will not be observed, namely the quantum coherence will be restored.
In fact, of all the possible effects of the fluctuations in spacetime topology
on the low energy physics, the most provocative one that immediately
attracted enormous excitement was the advocation initiated by Baum [1]
and by Hawking [1] and then refined later by Coleman [3] that the wormholes
have ultimately an effect of turning all the constants of nature into
``dynamical random variables''. Thus this Baum-Hawking-Coleman (BHC)
mechanism leads to a striking conclusion that the effects of 
(particularly) wormholes introduces into the low energy physics a
fundamental quantum indeterminancy of the values of the constants of
nature which can be thought of as an additional degree of uncertainty
over the usual uncertainty in quantum mechanics. Unlike the issue of
the loss of quantum coherence discussed earlier, however, this 
BHC-mechanism does not simply imply the elimination of the classical
predictability of nature by the effect of quantum gravity. For instance,
the BHC-mechanism actually leads to the prediction that the most
probable value of the fully-renormalized cosmological constant is
zero, indeed in exact agreement with the observation. For the detailed
arguments involved in the BHC-mechanism particularly concerning the
most probable value of the fully-renormalized cosmological constant,
we refer the reader to the literature [1,3]. But it seems fair to mention
that the formulation of wormhole physics particularly the one put
forward by Coleman is not without some inherent flaws. First, logically
the Coleman's wormhole physics formulation had faced some severe
criticisms such as ``sliding of Newton's constant problem'' [7] and
``large wormhole catastrophe'' [8] which seem to be unfortunately quite
generic. However, it seems that the most fundamental and crucial
difficulty associated with Coleman's formulation is the use of saddle
point approximation to the Euclidean path integral for quantum gravity
since, as is well-known, the Euclidean Einstein-Hilbert action is 
not bounded below [9]. \\
Now, then, note that both issues discussed thus far, namely the loss
of quantum coherence and the determination of probability distribution
for constants of nature (which are now random variables), are clearly
based on the assumption that there are wormhole instantons as saddle
points of the Euclidean action of the theory under consideration.
Therefore unless one can demonstrate that there are large class of
theories comprised of gravity with or without matter which admit
Planck-sized wormhole instantons as solutions to the classical field
equations, the discussion above on interesting effects of wormholes
on low energy physics will lose much of its meaning. Unfortunately,
thus far only a handful of restricted classes of theories are known
to possess classical wormhole instanton solutions and they include ;
Einstein-Kalb-Ramond (KR) antisymmetric tensor theory [4], Einstein-Yang-
Mills theory [5] and Einstein-complex scalar field theory in the presence
of spontaneous symmetry breaking [6]. The classical wormhole instanton
solutions and the semiclassical analysis of their effects on low
energy physics in these theories had been thoroughly studied in the
literature. Here in the present work, we revisit the 
Einstein-Yang-Mills (EYM) theory of Hosoya and Ogura [5] which is 
known to admit classical, Euclidean wormhole instanton solutions. 
One may wonder why anybody
should repeatedly go through a well-studied theory like this one.
Although the ``classical'' wormhole instanton as a solution to the
classical field equations and its effects on low energy physics had
been studied extensively, almost no attempt has been made concerning 
the serious study of ``quantum'' wormholes in the same theory.
Thus in the present work we attempt to explore the existence of
quantum wormholes in EYM theory. 
Now we describe briefly the approach we shall employ
to explore the physics of quantum wormholes in our theory.
And to do so, it seems necessary to distinguish between the definition
of ``classical'' wormhole and that of ``quantum'' wormhole.
In the classical sense, wormholes are Euclidean metrics which are
solutions to the Euclidean classical field equations representing
spacetimes consisting of two asymptotically Euclidean regions joined
by a narrow tube or throat. In the quantum regime, on the other hand,
and particularly in the context of the canonical quantum cosmology,
quantum wormholes may be identified
with a state or an excitation represented by a solution to the
Wheeler-DeWitt equation satisfying a certain boundary condition
describing the wormhole configuration. An widely-accepted such
``wormhole boundary condition'' is the one advocated by Hawking and
Page [10]. And it states that wormhole wave functions are supposed to
behave in such a way that they are damped, say, exponentially for
large 3-geometries ($\sqrt{h} \rightarrow \infty$) and are regular
in some suitable way when the 3-geometry collapses to zero
($\sqrt{h} \rightarrow 0$). Thus we shall construct a minisuperspace
quantum cosmology model possessing SO(4)-symmetry based on the 
EYM theory and attempt to solve associated Wheeler-DeWitt equation. 
It will be demonstrated that there indeed is a solution to the
Wheeler-DeWitt equation whose asymptotic behaviors
satisfy the ``wormhole boundary condition'' stated above and hence
can be identified with a wormhole wave function, namely a universe
wave function for quantum wormholes. \\
This paper is organized as follows : In sect.II, we recapitulate classical
wormhole instanton solutions and the semiclassical analysis of their
effects on low energy physics in EYM theory.
Sect.III will be devoted to the study of quantum 
wormholes in this theory employing the approach described above.
Finally in sect.IV, we summarize the results of our study and discuss
their physical implications.

\centerline {\rm\bf II. Classical wormhole physics in EYM theory}

We begin by reviewing the classical wormhole solution in EYM theory and
its implications on low energy physics in a rather different context from
the one in the literature [5].\\
Consider a system consisting of Yang-Mills (YM) theory of SU(2) gauge field
coupled to Einstein gravity in the presence of the cosmological constant.
This Einstein-Yang-Mills theory is represented by the 
Euclidean action [5]

\begin{eqnarray}
 I_{EYM} = \int_{M}d^{4}x\sqrt{g}
 [\Lambda -{{M_{p}^{2}}\over{16\pi}}R + {1\over 4g^2_{c}}F^{a}_{\mu\nu}
 F^{a\mu\nu}]
   - \int_{\partial{M}}d^{3}x\sqrt{h}{{M_{p}^{2}}\over{8\pi}}(K-K_{0}) 
\end{eqnarray}
where $F^{a}_{\mu\nu}$ is the field strength of the YM gauge field 
$A^{a}_{\mu}$ with $a = 1, 2, 3$ being the SU(2) group index and $g_{c}$
being the gauge coupling constant. 
We also added Gibbons-Hawking gravitational boundary term on $\partial M$ with
$h$ being the metric induced on $\partial M$ and $K$ being the trace of
the second fundamental form of $\partial M$. \\
Then by extremizing the action above with respect to the metric 
$g_{\mu\nu}$ and the YM gauge field $A^{a}_{\mu}$, 
one gets the following classical field equations respectively
\begin{eqnarray}
 &&R_{\mu\nu} - {1\over2}g_{\mu\nu}R + {{8\pi}\over{M^{2}_{p}}}\Lambda 
 g_{\mu\nu} = {{8\pi}\over{M^{2}_{p}}}T_{\mu\nu}  \\
 \rm{with}\quad
 &&T_{\mu\nu} = {1\over g^2_{c}}[F^{a}_{\mu\alpha}F_{\nu}^{a\alpha}
 - {1\over 4}g_{\mu\nu}(F^{a}_{\alpha\beta}F^{\alpha\beta})],\nonumber\\
 &&D_{\mu}[\sqrt{g}F^{a\mu\nu}] = 0
\end{eqnarray}
where and henceforth we use the notations
$F^{a}_{\mu\nu} = \partial_{\mu}A^{a}_{\nu} - \partial_{\nu}A^{a}_{\mu}
+ \epsilon^{abc}A^{b}_{\mu}A^{c}_{\nu}$,
$D^{ac}_{\mu} = \partial_{\mu}\delta^{ac} + \epsilon^{abc}A^{b}_{\mu}$,
$A_{\mu} = A^{a}_{\mu}(-iT^{a})$ and $F_{\mu\nu} = F^{a}_{\mu\nu}
(-iT^{a})$ with $D^{ac}_{\mu}$ being the usual gauge-covariant derivative
and $T^{a}$'s are SU(2) group generators obeying the SU(2) Lie algebra
$[T^{a}, T^{b}] = i\epsilon^{abc}T^{c}$ and the normalization
$Tr(T^{a}T^{b}) = \delta^{ab}/2$. \\
Now we look for a Euclidean, SO(4)-symmetric wormhole solution which is
an instanton describing the nucleation of a, presumably, Planck-sized
baby (spatially-closed ; $k = +1$) FRW-universe.
To this end, we begin by taking SO(4)-symmetric ans\H atz for the 
Euclidean metric and YM gauge field. First, the spatially-closed
FRW-metric, which has the SO(4)-symmetry, is given by
\begin{eqnarray}
 ds^{2} &=& g_{\mu\nu}dx^{\mu}dx^{\nu} =
            \delta_{AB} e^{A}{\otimes}e^{B} \\
 &=& \sigma^{2}[N^{2}(\tau)d{\tau}^{2} + 
 a^{2}(\tau)\sigma^{a}{\otimes}\sigma^{a}] \nonumber 
\end{eqnarray}
where $N(\tau)$ and $a(\tau)$ are lapse function and scale factor 
respectively and $\sigma^{2} = (2/3\pi M^2_{p})$ has been introduced
for convenience. Throughout this work, we shall work with
non-coordinate basis with indices $A, B = 0, a$ $(a = 1, 2, 3)$ rather
than with coordinate basis with indices 
$\mu, \nu = \tau, \theta, \phi, \psi$ where $(\theta, \phi, \psi)$
are Euler angles parametrizing the spatial section of the manifold
which has the geometry of $S^3$. Then the non-coordinate basis
1-forms can be read off as
\begin{eqnarray}
e^{A} = \{e^{0} = \sigma N(\tau)d\tau, ~~e^{a} = \sigma a(\tau)\sigma^{a}\}
\end{eqnarray}
where $\{\sigma^{a}\}$ $(a = 1, 2, 3)$ form a basis on the 3-sphere $S^3$
as just mentioned satisfying the SU(2) ``Maurer-Cartan'' structure 
equation
\begin{eqnarray}
 d\sigma^{a} = {1\over2}\epsilon^{abc}\sigma^{b}{\wedge}\sigma^{c}
\end{eqnarray}
and can be represented in terms of 3-Euler angles $0\le\theta\le\pi$,
$0\le\phi\le2\pi$ and  $0\le\psi\le4\pi$, parametrizing $S^{3}$

\begin{eqnarray}
 \sigma^{1} &=& cos{\psi}d{\theta} + sin{\psi}sin{\theta}d{\phi},\nonumber\\
 \sigma^{2} &=& sin{\psi}d{\theta} - cos{\psi}sin{\theta}d{\phi}, \\
 \sigma^{3} &=& d{\psi} + cos{\theta}d{\phi}.\nonumber
\end{eqnarray}
For later use, we also write down the associated vierbein and its inverse 
using the definition, $e^{A} = e^{A}_{\mu}dx^{\mu}$ ,
$e^{A}_{\mu}e^{\mu}_{B} = \delta^{A}_{B}$ and 
$e^{\mu}_{A}e^{A}_{\nu} = \delta^{\mu}_{\nu}$ where
$x^{\mu} = (\tau,\theta,\phi,\psi)$.
\begin{eqnarray}
 e^{A}_{\mu} = \sigma \left(\matrix
               { N & 0          & 0                     & 0 \cr
                 0 & acos{\psi} & asin{\psi}sin{\theta} & 0 \cr
                 0 & asin{\psi} &-acos{\psi}sin{\theta} & 0 \cr
                 0 & 0          & acos{\theta}          & a }
               \right)\quad,\quad
 e_{A}^{\mu} = {1\over \sigma} \left(\matrix
      { {1\over N} & 0          & 0                     & 0 \cr
                 0 & {1\over{a}}cos{\psi}
                   & {1\over{a}}sin{\psi}
                   & 0 \cr
                 0 & {sin{\psi}\over{asin{\theta}}}
                   & {-cos{\psi}\over{asin{\theta}}}
                   & 0 \cr
                 0 & {-sin{\psi}cos{\theta}\over{asin{\theta}}}
                   & { cos{\psi}cos{\theta}\over{asin{\theta}}}
                   & {1\over{a}} }
               \right).
\end{eqnarray}
Next we turn to the choice of SO(4)-symmetric ans\H atz for the YM
gauge field and its field strength. Note that the SU(2) group
manifold is also $S^3$ just like it is the case for the geometry of
the spatial section of the spacetime manifold. Thus one may choose
the left-invariant 1-form $\{\sigma^{a}\}$ as the ``common'' basis
for both group manifold and the spatial section of the spacetime
manifold. And this indicates that there is now ``mixing'' between
the group index in the YM field and the non-coordinate basis frame
index since we choose to work with non-coordinate basis.
In the present work, we restrict our interest only to the 
magnetically-charged solution, i.e., $F^{a}_{\tau i} = 0$
$(i = \theta, ~\phi, ~\psi)$. The motivation for this restriction
will become clearer later on in the discussion of quantum wormhole
physics. Namely from
$F^{a}_{\mu\nu} = \{F^{a}_{\tau i}=0, ~F^{a}_{ij}\}$
$(i, j = \theta, ~\phi, ~\psi)$, we can write down the SO(4)-symmetric
ans\H atz for YM field strength in non-coordinate basis indices
\begin{eqnarray}
F^{a}_{0b} &=& e^{\mu}_{0}e^{\nu}_{b}F^{a}_{\mu\nu}
            =  e^{\tau}_{0}e^{i}_{b}F^{a}_{\tau i} = 0, \\
F^{a}_{bc} &=& e^{\mu}_{b}e^{\nu}_{c}F^{a}_{\mu\nu}
            =  e^{i}_{b}e^{j}_{c}F^{a}_{ij} = f(\tau)\epsilon^{abc}. 
            \nonumber
\end{eqnarray}
Note that the choice of the ans\H atz for $F^{a}_{bc}$ above is
motivated by the choice of ans\H atz for the Kalb-Ramond antisymmetric
tensor field strength, $H_{\mu\nu\lambda} = h(\tau)\epsilon_{\mu\nu\lambda}$
in the work by Giddings and Strominger [4] who looked for SO(4)-symmetric
wormhole instanton solutions as well. As a matter of fact, this is an
essential common structure shared by the Einstein-antisymmetric tensor
theory considered by Giddings and Strominger [4] and the present EYM theory
first studied by Hosoya and Ogura [5] that allows the existence of classical
wormhole solution. Also note that this kind of starting point of ours is
{\it different} from that in the original work of Hosoya and Ogura [5] in 
which they begin by taking the SO(4)-symmetric ans\H atz for the YM
gauge potential rather than its field strength as we are doing here.
The motivation for taking the ans\H atz for the field strength
$F^{a}_{AB}$ will become clearer shortly. \\
Now, we would like to determine the condition that the function $f(\tau)$
appearing in $F^{a}_{bc}$ should meet and the form of the associated YM
gauge potential $A^{a}_{B}$ by demanding only that they be consistent 
with the definition $F^{a} = dA^{a} + {1\over 2}\epsilon^{abc}
A^{b}\wedge A^{c}$ and obey the Bianchi identity
$DF^{a} = dF^{a} - \epsilon^{abc}F^{b}\wedge A^{c} = 0$.
And to do so, we choose to take the temporal gauge $A^{a}_{\tau} = 0$
(or equivalently $A^{a}_{0} = 0$ in non-coordinate basis). This gauge
choice is indeed natural since the spacetime is homogeneous and isotropic
thus depends only on $\tau $-coordinate, there is no gauge freedom
associated with the space-dependent gauge transformation. Now we begin
with the definition of the YM field strength $F^{a} = dA^{a} + {1\over 2}
\epsilon^{abc} A^{b}\wedge A^{c}$. On the one hand, using the YM gauge
potential expressed in non-coordinate basis (in $A^{a}_{\tau} = 0$ gauge)
$A^{a} = A^{a}_{\mu}dx^{\mu} = A^{a}_{b}e^{b}$ $(b = 1, 2, 3)$, one can
evaluate the above defining equation for the field strength with the help
of Maurer-Cartan structure equation $d\sigma^{a} = {1\over 2}
\epsilon^{abc} \sigma^{b}\wedge \sigma^{c}$ given earlier. On the other
hand, from our choice of SO(4)-symmetric ans\H atz for the YM field 
strength, one has $F^{a} = {1\over 2}F^{a}_{\mu\nu}dx^{\mu}\wedge
dx^{\nu} = {1\over 2}F^{a}_{bc}(e^{b}\wedge e^{c}) = {1\over 2}\sigma^2
a^{2}(\tau)f(\tau)\epsilon^{abc} (e^{b}\wedge e^{c})$. Then by equating
these two alternative expressions for the field strength 2-form $F^{a}$,
one gets a set of two equations
\begin{eqnarray}
\partial_{\tau}[a&(\tau)& A^{a}_{b}(\tau)] = 0, \\
A^{a}_{b}(\tau)\epsilon^{bde} &+& \sigma a(\tau)A^{b}_{d}(\tau)
A^{c}_{e}(\tau)\epsilon^{abc} = \sigma a(\tau)f(\tau) \epsilon^{ade}.
\nonumber 
\end{eqnarray}
Immediately, one can realize that these two equation can be simultaneously
satisfied provided
\begin{eqnarray}
A^{a}_{b}(\tau) &=& {k\over \sigma a(\tau)}\delta^{a}_{b}, \\
f(\tau) &=& {k(k + 1) \over \sigma^2 a^2(\tau)}. \nonumber
\end{eqnarray}
Our next job would then be to demonstrate that this choice of the
SO(4)-symmetric ans\H atz for YM gauge potential  and its field strength
tensor are legitimate since they indeed satisfy the Bianchi identity.
As a matter of fact, it is straightforward to check that the Bianchi
identity $DF^{a} = dF^{a} - \epsilon^{abc}F^{b} \wedge A^{c} = 0$ is
satisfied with our choice of ans\H atz, $A^{a} = k\sigma^{a}$ and
$F^{a} = {1\over 2}k(k + 1)\epsilon^{abc}(\sigma^{b} \wedge \sigma^{c})$.
Therefore, we can conclude that a consistent choice of the SO(4)-symmetric
ans\H atz for non-vanishing components of YM gauge potential and its 
field strength are
\begin{eqnarray}
A^{a}_{b}(\tau) &=& e^{\mu}_{b}A^{a}_{\mu} = {k\over \sigma a(\tau)}
\delta^{a}_{b}, \\
F^{a}_{bc}(\tau) &=& e^{\mu}_{b}e^{\nu}_{c}F^{a}_{\mu\nu} =
{k(k + 1)\over \sigma^2 a^2(\tau)}\epsilon^{abc}. \nonumber
\end{eqnarray}
Notice, again, that in contrast to the formulation employed in the original
work by Hosoya and Ogura [5] where they obtained these types of expressions for
YM gauge potential and its field strength tensor only after explicitly
solving the classical YM field equation, here we obtained them simply
from definitions and the Bianchi identity. Therefore, we particularly
would like to stress that these SO(4)-symmetric ans\H atz for $A^{a}$
and $F^{a}$ remain valid even off-shell as well as on-shell. In other words,
these expressions for the YM gauge potential and its field strength can
be used for both classical and quantum treatment that we shall discuss
later on. As announced in the introduction, the aim of this work is to
study the physics of classical and quantum wormholes in EYM theory. And we
are supposed to study the classical wormhole solutions and their effects
on low-energy physics in the present section and the quantum wormholes
later on. To repeat, wormholes are Euclidean metrics which are solutions
to the Euclidean classical field equations representing spacetimes
consisting of two asymptotic Euclidean regions joined by a narrow tube
or throat. Thus to study classical wormholes, we need to first look for
exact solutions to the Euclidean classical field equations which exhibit
above-mentioned geometrical structure. Note that we already have chosen
SO(4)-symmetric ans\H atz for the YM gauge potential and its field 
strength which are consistent with the definitions and the Bianchi
identity. Thus before we attempt to solve the Einstein equation, we
need to check if these ans\H atz for YM fields can be solution to
the classical YM field equation. As we shall see shortly, demanding that
they be a solution amounts to fixing the parameter ``$k$'' appearing in the
ans\H atz. Now, since we choose to work in non-coordinate basis, we also
consider the classical YM field equation in non-coordinate basis as well.
And in non-coordinate basis, the YM field equation is obtained by simply
replacing $F^{a}$ with its Hodge dual $\tilde{F}^{a}$ in the Bianchi
identity equation given earlier, namely
\begin{eqnarray}
D\tilde{F}^{a} = d\tilde{F}^{a} - \epsilon^{abc}\tilde{F}^{b}
\wedge A^{c} = 0. 
\end{eqnarray}
Since the Hodge dual of the YM field strength is obtained to be
\begin{eqnarray}
\tilde{F}^{a} = k(k + 1){N(\tau) \over a(\tau)}(d\tau \wedge 
\sigma^{a}),
\end{eqnarray}
it is again straightforward to confirm that our choice of the SO(4)-
symmetric YM gauge field given earlier does indeed satisfy the classical
YM field equation provided $k = - 1/2$, namely if
\begin{eqnarray}
A^{a} &=& k\sigma^{a} = -{1\over 2}\sigma^{a}, \\
F^{a} &=& {1\over 2}k(k + 1)\epsilon^{abc}(\sigma^{b} \wedge \sigma^{c})
= -{1\over 8}\epsilon^{abc}(\sigma^{b} \wedge \sigma^{c}). \nonumber
\end{eqnarray}
Next, we turn to the Einstein field equation. Since the SO(4)-symmetric
ans\H atz for YM fields are given in terms of the scale factor $a(\tau)$,
the coupled Einstein-YM field equations essentially reduce to the Einstein
field equation of the scale factor $a(\tau)$ alone. \\
Now consider the $\tau \tau$-component of the Einstein field equation
\begin{eqnarray}
R_{\tau \tau} - {1\over 2}g_{\tau \tau}R + {8\pi \over M^2_p}\Lambda
g_{\tau \tau} = {8\pi \over M^2_p}T_{\tau \tau}.
\end{eqnarray}
The YM field energy-momentum tensor in coordinate basis, $T_{\tau \tau}$,
is related to its counterpart in non-coordinate basis by
$T_{\tau \tau} = e^{A}_{\tau} e^{B}_{\tau} T_{AB} = 
e^{0}_{\tau} e^{0}_{\tau} T_{00}$ and $T_{00}$ can be easily computed by
YM field strength in non-coordinate basis, $F^{a}_{0b} = 0$,
$F^{a}_{bc} = (-{1\over 4}){1\over \sigma^2 a^2(\tau)}\epsilon^{abc}$ to
be $T_{00} = (-{3\over 32}){1\over \sigma^4 g^2_{c} a^{4}(\tau)}$.
Thus also using,
\begin{eqnarray}
R_{\tau \tau} &=& [-3({\ddot{a}\over a}) + 3({\dot{N}\over N})
({\dot{a}\over a})], \nonumber \\
R &=& {6\over \sigma^2}[({\dot{N}\over N^3})({\dot{a}\over a}) -
{1\over N^2}({\ddot{a}\over a} + ({\dot{a}\over a})^2) + {1\over a^2}]
\nonumber
\end{eqnarray}
where overdot denotes the derivative with respect to the Euclidean time
$\tau$ and $\sigma^2 \equiv 2/3\pi M^2_{p}$, one gets the $\tau \tau$-
component of the Einstein equation as 
\begin{eqnarray}
({da\over d\tau})^2 = N^2 - \lambda N^2 a^2 - N^2 {r^2_{0}\over a^2}
\end{eqnarray}
where $r^2_{0} \equiv 3\pi^2/8g^2_{c}$ and we introduced ``dimensionless
cosmological constant'' as $\lambda \equiv 16\Lambda/9M^4_{p}$.
Now, we attempt to solve this Einstein field equation with different gauge
choice for the lapse $N(\tau)$ and in the presence/absence of the 
cosmological constant. \\
{\bf 1. In the absence of the cosmological constant} \\
(i) With the gauge choice $N(\tau) = 1$ : \\
The Einstein equation above reduces to
\begin{eqnarray}
({da\over d\tau})^2 = 1 - {r^2_{0}\over a^2}
\end{eqnarray}
of which the solution is given by
\begin{eqnarray}
\tau = \int_{0}^{\tau} d\tau' = \int_{a(0)=r_{0}}^{a} {ada\over
\sqrt{a^2 - r^2_{0}}} \nonumber
\end{eqnarray}
which yields upon the integration
\begin{eqnarray}
a(\tau) = [r^2_{0} + \tau^2]^{1/2}
\end{eqnarray}
where we chose that the wormhole throat, i.e., the minimum value of 
$a(\tau)$ occurs for $\tau = 0$, namely
${da\over d\tau}\mid_{\tau=0} = 0$ and $a(\tau=0) = r_{0}$. This
solution represents a Tolman-type wormhole found first by Hosoya
and Ogura [5]. \\
(ii) With the gauge choice $N(\tau) = a(\tau)$ : \\
The Einstein equation, this time, reduces to
\begin{eqnarray}
({da\over d\tau})^2 = a^2 - r^2_{0}
\end{eqnarray}
of which the solution is given by
\begin{eqnarray}
\tau = \int_{0}^{\tau} d\tau' = \int_{a(0)=r_{0}}^{a} {da\over
\sqrt{a^2 - r^2_{0}}} \nonumber
\end{eqnarray}
which yields upon performing the integration
\begin{eqnarray}
a(\tau) = r_{0} \cosh (2\tau).
\end{eqnarray}
{\bf 2. In the presence of the cosmological constant} \\
(i) With the gauge choice $N(\tau) = 1$ : \\
The Einstein equation reads
\begin{eqnarray}
({da\over d\tau})^2 = 1  - \lambda a^2 - {r^2_{0}\over a^2}.
\end{eqnarray}
As usual, we choose the minimum value of $a(\tau)$, i.e., the wormhole
neck to occur for $\tau = 0$. Then from
\begin{eqnarray}
({da\over d\tau})^2\mid_{\tau=0} = {1\over a^2}[- \lambda a^4 + a^2 - r^2_{0}]
= 0 \nonumber
\end{eqnarray}
which has zeros at 
$r^2_{\pm} = {1\over 2\lambda}[1 \pm \sqrt{1 - 4\lambda r^2_{0}}]$,
the radius of the wormhole neck, which exists when $\lambda < 1/4r^2_{0}$,
is found to be
\begin{eqnarray}
a(0) = r_{-} = [{1\over 2\lambda}(1 - \sqrt{1 - 4\lambda r^2_{0}})]^{1/2}.
\end{eqnarray}
Thus, upon carrying out the integration
\begin{eqnarray}
\tau = \int_{0}^{\tau} d\tau' = \int_{r_{-}}^{a} {ada\over
\sqrt{-\lambda a^4+a^2-r^2_{0}}}, \nonumber
\end{eqnarray}
one gets
\begin{eqnarray}
a(\tau) = [{1\over 2\lambda}\{1 - \sqrt{1 - 4\lambda r^2_{0}}\cos (2
\sqrt{\lambda}\tau)\}]^{1/2}.
\end{eqnarray}
This solution for the scale factor $a(\tau)$ takes the minimum value at
$\tau = 0$, $a_{min} = [{1\over 2\lambda}(1 - \sqrt{1 - 
4\lambda r^2_{0}})]^{1/2}$ and the maximum value at 
$\tau = \pi/2\sqrt{\lambda}$,
$a_{max} = [{1\over 2\lambda}(1 + \sqrt{1 - 4\lambda r^2_{0}})]^{1/2}$.
Thus this solution represents a typical wormhole configuration. \\
(ii) With the gauge choice $N(\tau) = a(\tau)$ : \\
In this gauge, the Einstein equation reads
\begin{eqnarray}
({da\over d\tau})^2 = a^2 - \lambda a^4 - r^2_{0}.
\end{eqnarray}
Once again, we choose the minimum value of $a(\tau)$, i.e., the wormhole neck
to occur for $\tau = 0$, i.e., $({da\over d\tau})\mid_{\tau=0} = 0$, then
$a(0) = r_{-} = [{1\over 2\lambda}(1 - \sqrt{1 - 4\lambda r^2_{0}})]^{1/2}$
which is possible provided $\lambda < 1/4r^2_{0}$. Then using
$(-\lambda a^4+a^2-r^2_{0}) = -\lambda (a^2 - r^2_{+})(a^2 - r^2_{-})$,
\begin{eqnarray}
\tau = \int_{0}^{\tau} d\tau' &=& \int_{r_{-}}^{a} {da\over
\sqrt{-\lambda a^4+a^2-r^2_{0}}} \nonumber \\
&=& \int_{r_{-}}^{a} {da\over \sqrt{-\lambda (a^2 - r^2_{+})(a^2 - r^2_{-})}}.
\nonumber
\end{eqnarray}
This integral can be expressed in terms of the elliptic integral and it is
\begin{eqnarray}
\sqrt{-\lambda}r_{+}\tau = F({a\over r_{-}}, ~{r_{-}\over r_{+}}) -
F(1, ~{r_{-}\over r_{+}})
\end{eqnarray}
where $F(x, k)$ denotes the elliptic integral of the 1st kind $(0<k<1)$ and
$F(x=1, k) \equiv K(k)$ is the ``complete elliptic integral'' of the 1st kind.
\\
Having constructed wormhole instanton solutions with different gauge choices 
for the lapse $N(\tau)$, we now turn to their contribution to the topology-
changing tunnelling amplitude. Namely, we evaluate the wormhole instanton 
action $I_{EYM}(instanton)$ (here $I_{EYM}$ denotes the Euclidean action of
the EYM theory), i.e., the minimum Euclidean action of the wormhole instanton
configuration which makes dominant contribution to the topology-changing
tunnelling amplitude. And this amounts to substituting the Einstein field
equation (its trace) satisfied by the wormhole instanton solution
\begin{eqnarray}
R = {32\pi\over M^2_{p}}\Lambda \nonumber
\end{eqnarray}
into the Euclidean EYM theory action given earlier to obtain
\begin{eqnarray}
I_{EYM}(instanton) = \int d^4x\sqrt{g}[- \Lambda + {1\over 4g^2_{c}}
(F^{a}_{\mu\nu})^2].
\end{eqnarray}
Then using, $\int d^4x\sqrt{g} = (2\pi^2 \sigma^4)\int_{0}^{\infty} d\tau
Na^3$, $\sigma^2 = 2/3\pi M^2_{p}$, $\lambda = 16\Lambda/9M^4_{p}$
and $(F^{a}_{\mu\nu})^2 = (F^{a}_{bc})^2 = 3/8\sigma^4 a^4$, this
becomes
\begin{eqnarray}
I_{EYM}(instanton) &=& \int_{0}^{\infty}d\tau Na^3 [- {8\Lambda\over 9M^4_{p}}
+ {3\pi^2\over 16g^2_{c}}{1\over a^4}] \\
&=& {1\over 2}\int_{0}^{\infty}d\tau ({N\over a})[- \lambda a^4 + r^2_{0}]
= {1\over 2}\int_{R_{-}}^{a(\infty)} da {(-\lambda a^4 + r^2_{0})\over
\sqrt{-\lambda a^4 + a^2 - r^2_{0}}} \nonumber
\end{eqnarray}
where we used the $\tau \tau$-component of the Einstein equation
$(da/d\tau)^2 = N^2(1-\lambda a^2-r^2_{0}/a^2)$ which yields
$d\tau (N/a) = da/\sqrt{-\lambda a^4+a^2-r^2_{0}}$. Note, first, that above
expression for $I_{EYM}(instanton)$ is completely independent of the lapse 
function $N(\tau)$. Since the relevant gauge freedom for the case at hand
is the arbitrariness in choosing the lapse $N(\tau)$, this means that the
wormhole instanton action $I_{EYM}(instanton)$ has gauge-invariance. As a
matter of fact, since the quantity $\exp{[-I_{EYM}(instanton)]}$ represents
the semi-classical approximation to the topology-changing tunnelling
amplitude, it is a physical observable which should have manifest gauge-
invariance. And we have just confirmed this point. \\
Next note that unfortunately, it is not possible to carry out explicitly
the integral in the expression for $I_{EYM}(instanton)$ above to obtain
the precise value of the wormhole instanton action. As was pointed out by 
Hosoya and Ogura [5], however, for a sufficiently small cosmological constant,
this instanton action may be expanded in powers of the cosmological
constant $\lambda = 16\Lambda/9M^4_{p}$ to yield
\begin{eqnarray}
I_{EYM}(instanton) = - {1\over 3\lambda} - {3\pi\over 4\alpha_{g}}
\ln (r_{0}\sqrt{\lambda}) + O(\sqrt{\lambda})
\end{eqnarray}
where $r^2_{0} = 3\pi^2/8g^2_{c}$ as introduced earlier and
$\alpha_{g} \equiv g^2_{c}/4\pi$. The first term results from the
contribution from the half of the de Sitter instanton while the second term
represents the contribution from the wormhole (neck) geometry especially
with the logarithmic factor indicating the infrared cut-off generated by
the finite wormhole neck size. We now end with few remarks. Firstly, note
that the expression for $I_{EYM}(instanton)$ given above in eq.(29) is valid
only for sufficiently small values of the cosmological constant. Next, 
consider the contribution from the wormhole configuration to the topology-
changing tunnelling amplitude 
\begin{eqnarray}
\exp {[{3\pi\over 4\alpha_{g}}\ln (r_{0}\sqrt{\lambda})]}. \nonumber
\end{eqnarray}
As Hosoya and Ogura [5] remarked, at first sight, this factor appears to serve
as a suppression rather than an enhancement for the tunnelling amplitude
since $\lambda$ is assumed to be small anyway. Besides, wormholes with
larger size, namely, larger than the scale set by the dimensionless
cosmological constant, i.e., $r_{0} >> 1/\sqrt{\lambda}$ appears to enhance the
tunnelling amplitude than the ones with smaller size. This last point is
particularly disappointing since it is just against our expectation. 
As a matter of fact, it is noteworthy that just the opposite happens in the
physics of axionic wormholes studied by Giddings and Strominger [4] where the
tunnelling amplitude is highly suppressed for wormholes with size greater
than the Planck length. All these arguments might be rather hasty since it
is the whole quantity
\begin{eqnarray}
\exp{[-I_{EYM}(instanton)]} = e^{1\over 3\lambda}
(r_{0}\sqrt{\lambda})^{3\pi\over 4\alpha_{g}}
\end{eqnarray}
that one really should consider as the right tunnelling amplitude. And for
sufficiently small $\lambda$, the first singular factor dominates over the
second factor (coming from the wormhole contribution) still producing a
sizable tunnelling amplitude. Neverthless, since it is the role played by
the small cosmological constant that essentially elevates the tunnelling 
amplitude to a sizable magnitude, it seems fair to say that the wormhole
instanton solutions in this EYM theory do not seem to be local minima of
the Euclidean action which have manifest contribution to the semiclassical
tunnelling amplitude. 

\centerline {\rm \bf III. Quantum Wormholes in Einstein-Yang-Mills Theory} 

Thus far, we have considered classical treatment of wormholes in EYM theory.
Namely we constructed SO(4)-symmetric metric solutions to the classical
EYM field equations that exhibit the geometry of wormhole spacetimes and
studied their nature. We now turn to the quantum treatment of wormholes
in the same EYM theory. The formulation we shall employ can be 
summarized as follows : we construct and study a minisuperspace model
(again based on SO(4)-symmetry) of canonical quantum cosmology in which the
main objective is to solve the Wheeler-DeWitt (WD) equation to find the
universe wave function. And to see if there is an excitation that can be
interpreted as a ``quantum wormhole'', we look for a particular solution
to the WD equation with ``wormhole boundary condition''. Here the wormhole
boundary condition refers to an appropriate boundary condition of a universe
wave function that allows one to naturally identify the universe wave
function as representing an excitation corresponding to a wormhole state.
And as a proposal for such wormhole boundary condition, we shall employ the 
one advocated by Hawking and Page. According to their proposal, wormhole 
wave functions are supposed to behave in such a way that they are damped,
say, exponentially, for large 3-geometries ($\sqrt{h} \rightarrow \infty$)
and are regular in some suitable way when the 3-geometry collapses to zero
($\sqrt{h} \rightarrow 0$). We now begin by constructing a minisuperspace
quantum cosmology model based on EYM theory described by the action given
earlier. As already stated, to do so, we choose to take the avenue of
canonical quantum cosmology based on Arnowitt-Deser-Misner (ADM)'s (3+1)-
space-plus-time split formalism [11-13]. As usual, then, in order to render the
system tractable, we reduce the infinite-dimensional ``superspace'' down to
a finite-dimensional minisuperspace by assuming that the 4-dim. spacetime
has the geometry of spatially-closed ($k = +1$) FRW-metric which is the one
we adopted in the classical treatment of the system earlier. The geometry
of its spatial section is, then, that of  $S^3$ and hence it posseses
SO(4)-symmetry. Now, since the spacetime geometry is taken to possess the
SO(4)-symmtry, the matter field defined on it, i.e., the YM gauge field
should have the same SO(4)-symmetry. Thus the SO(4)-symmetric ans\H atz 
for the metric is again the spatially-closed FRW-metric
\begin{eqnarray}
ds^2 &=& \sigma^2 [N^2(\tau)d\tau^2 + a^2(\tau)\sigma^{a}\otimes \sigma^{a}]
\nonumber \\
&=& \eta_{AB}e^{A}\otimes e^{B} 
\end{eqnarray}
and the SO(4)-symmetric ans\H atz for the YM gauge potential and its field
strength tensor are, in non-coordinate basis
\begin{eqnarray}
A^{a}_{0}(\tau) &=& 0, ~~~A^{a}_{b}(\tau) = {k\over \sigma a(\tau)}
\delta^{a}_{b} \\
F^{a}_{0b}(\tau) &=& 0, ~~~F^{a}_{bc}(\tau) = {k(k+1)\over \sigma^2 a^2(\tau)}
\epsilon^{abc} \nonumber
\end{eqnarray}
where $A,B = 0,1,2,3$ and $a,b,c = 1,2,3$ and again $\sigma^2 = 2/3\pi M^2_{p}$.
Note first that in the metric, if $N$ is imaginary, it is the Lorentzian metric
with Lorentzian time $\tau = t$ and if $N$ real, it is the Euclidean metric with
Euclidean time $\tau$. As is well-known and as we shall see later, the WD equation
and its solution are independent of $N$ and $\tau$ and hence remain the same in
both the Lorentzian and Euclidean signatures. Next, recall that the SO(4)-symmetric
ans\H atz for $A^{a}_{B}(\tau)$ and $F^{a}_{BC}(\tau)$ given above remain valid
off-shell as well as on-shell as we stressed in the classical treatment of the
system discussed earlier. And it is because we obtained them simply from definitions
and the Bianchi identity which are bottomline conditions that should be met in
quantum formulations as well. Thus in the present quantum treatment, the parameter
$k$ appearing in the ans\H atz for YM gauge field are undetermined. (As we have
seen, it gets determined to be $k = -1/2$ only when one imposes on-shell condition,
namely only if one demands the classical YM field equation to be satisfied.)
Now, for the sake of definiteness, we choose the Euclidean signature (i.e., 
$N(\tau)$ is real) and write the Euclidean action of the EYM theory given earlier
in terms of these SO(4)-symmetric ans\H atz for the metric and YM gauge field.
Namely, using
$\int{d}^{4}x\sqrt{g}  = {\int}d\tau N(\int_{S^{3}}d^{3}x\sqrt{h}) 
 = (2{\pi}^{2}\sigma^{4}){\int}d\tau Na^{3}$ and 
\begin{eqnarray}
R &=& {6\over \sigma^2}[({\dot{N}\over N^3})({\dot{a}\over a}) -
{1\over N^2}({\ddot{a}\over a} + ({\dot{a}\over a})^2) + {1\over a^2}], 
\nonumber \\
(F^{a}_{\mu\nu})^2 &=& (F^{a}_{bc})^2 = 6 {k^2(k+1)^2 \over \sigma^4 a^4(\tau)}
\end{eqnarray}
with the overdot denoting the derivative with respect to $\tau$, the Euclidean
action of EYM theory takes the form
\begin{eqnarray}
 I_{G} &=& {\int}d^{4}x\sqrt{g}[\Lambda - {M^{2}_{p}\over16\pi}R]
       = {1\over2}{\int}d\tau Na^{3}[\lambda - \{{1\over{a^{2}}} 
       + ({{\dot a}\over{Na}})^{2}\}], \nonumber \\
 I_{YM} &=& {\int}d^{4}x\sqrt{g}[{1\over 4g^2_{c}}(F^{a}_{\mu\nu})^2]
         = (2{\pi}^{2}\sigma^{4}){\int}d\tau Na^{3}[{1\over 4g^2_{c}}
            \{6 {k^2(k+1)^2 \over \sigma^4 a^4}\}] \nonumber \\
        &=& {1\over2}{\int}d\tau Na^{3}[{r^2_{0}\over a^4}]  
\end{eqnarray} 
and hence
\begin{eqnarray}
 I_{EYM} &=& I_{G} + I_{YM}  \\
 &=& {1\over2}{\int}d\tau Na^{3}
     [\lambda - \{{1\over{a^{2}}} + ({{\dot a}\over{Na}})^{2}\} + 
     {r^2_{0}\over a^4}]. \nonumber
\end{eqnarray}
For the sake of completeness we also provide the Lorentzian action of the
EYM theory which can be obtained via $iS_{EYM} = - I_{EYM}$ and $\tau = it$
\begin{eqnarray}
S_{EYM} = {1\over2}{\int}dt Na^{3}
     [-\lambda + \{{1\over{a^{2}}} - ({{\dot a}\over{Na}})^{2}\} -
     {r^2_{0}\over a^4}] \equiv \int dt L_{ADM}
\end{eqnarray}
where now the overdot denotes the derivative with respect to the Lorentzian
time $t$ and again $\lambda$ denotes the dimensionless cosmological constant
$\lambda = 16\Lambda / 9M^4_{p}$ and $r^2_{0} \equiv {k^2(k+1)^2}6\pi^2 /g^2_{c}$
which becomes its classical counterpart $r^2_{0} = 3\pi^2 / 8g^2_{c}$ for 
$k = -1/2$. Now the subsequent procedure toward the derivation of the WD
equation via the Legendre transformation to the classical Hamiltonian and the
Dirac quantization can be best demonstrated in familiar Lorentzian signature.
Thus from this point on we work with the Lorentzian action for the construction
of the WD equation. Before we proceed, however, here we make one important remark
on the structure of the EYM theory action written in SO(4)-symmetric ans\H atz
given above. The fact that we choose to consider only the magnetically-charged 
case ($F^{a}_{0b} = 0$) and the fact that the SO(4)-symmetric ans\H atz for YM
field strength is written as $F^{a}_{bc} = {k(k+1)\over \sigma^2 a^2}
\epsilon^{abc}$ effectively reduce the EYM theory action to the action of a
single minisuperspace variable, i.e., the scale factor $a(\tau)$ alone. Thus
the resulting WD equation will also become a one-dim. Schr\H odinger-type
equation for the universe wave function $\Psi [a]$. As we shall see shortly,
this reduction to one-dim. problem of minisuperspace model essentially makes
it possible to find the universe wave function which represents a quantum
wormhole state. Formally speaking, this reduction is achieved by the full
gauge-fixing which consists of the gauge choice $N^{i}=0$ associated with
the 3-dim. diffeomorphism in the gravity sector and the gauge choice
$A^{a}_{0} = 0$ in the YM sector. \\
Now getting back to the derivation procedure for the WD equation, we
identify the canonical momentum conjugate to the minisuperspace variable $a$
with
\begin{eqnarray}
p_{a} = {\partial L_{ADM} \over \partial \dot{a}} = {a\over N}(-\dot{a})
\end{eqnarray}
and then obtain the classical Hamiltonian via the Legendre transformation
\begin{eqnarray}
 S_{EYM} &=& {\int}dtL_{ADM} \\
 &=& {\int}dt(p_{a}{\dot a} - H_{ADM})
 = {\int}[p_{a}da - (NH_{0} + N_{i}H^{i})dt]\nonumber
\end{eqnarray}
where $N_{i} = 0$ and $H_{ADM} = NH_{0} + N_{i}H^{i}$ denotes the ADM
Hamiltonian consisting of the sum of secondary constraints $H_{0} = 0$
and $H^{i} = 0$ as we shall discuss shortly. Namely the classical
Hamiltonian is obtained as
\begin{eqnarray}
 {{\delta}S_{EYM}\over{\delta}N}
 &=& {1\over2}a^{3}[-\lambda  + {1\over{a^{2}}} 
     + {{\dot a}^{2}\over{N^{2}a^{2}}} - {r^2_{0}\over a^4}] \\
 &=& {1\over{2a}}[p^{2}_{a} - \lambda a^{4} + a^{2} - r^2_{0}]. \nonumber 
\end{eqnarray}
Namely,    
\begin{eqnarray}
 H_{0} = -{{\delta}S_{EYM}\over{\delta}N}
 &=& {1\over{2a}}[-p^{2}_{a} 
     + (\lambda a^{4} - a^{2} + r^2_{0})]\nonumber\\
 &\equiv& {1\over{2a}}[-p^{2}_{a} + U(a)] \\
 {\rm where}\quad U(a) &=& {\lambda}a^{4} - a^{2} + r^2_{0}.
 \nonumber
\end{eqnarray}
General relativity is one of the most well-known constrained system.  
The invariance of the system under the 4-dim. diffeomorphisms (consisting of 
the time-reparametrization and the 3-dim. general coordinate transformations 
of the spacelike hypersurface) leads to the emergence of 4-constraint 
equations.  Of them, we need not explicitly impose the 3-momentum constraint 
equations since we already have taken the $N^{i} = 0$ gauge which amounts to 
assuming the $SO(4)$-symmetric spatially-closed FRW-metric.  
Thus, we only need to impose the 
Hamitonian constraint $H_{0} = 0$.  The classical Hamiltonian constraint now 
reads

\begin{eqnarray}
 H_{0} 
 = {1\over{2a}}[-p^{2}_{a} 
 + ({\lambda}a^{4} - a^{2} + r^2_{0})] = 0.
\end{eqnarray}
Next, in order to quantize this EYM system, we 
need to turn to the ``Dirac quantization procedure'' for the constrained 
system.  According to the Dirac quantization procedure, the invariance in the 
action of the theory under the 4-dim. diffeomorphisms is secured by demanding 
that the physical (universe) wave function $\Psi $ be annihilated by 
``operator versions'' of the 4-constraints.  Therefore, the classical 
Hamitonian constraint above turns into its quantum version, namely the 
WD equation given by

\begin{eqnarray}
 {\hat H_{0}}(p_{a} = -i{{\partial}\over{\partial}a})\Psi[a] = 0.
\end{eqnarray}
In order to obtain the correct form of this WD equation, we first 
examine the structure of the classical Hamiltonian

\begin{eqnarray}
 H_{0} 
 &=& {1\over{2a}}[-p^{2}_{a} 
 + ({\lambda}a^{4} - a^{2} + r^2_{0})] \nonumber \\
 &=& T + V \equiv {1\over2}G^{\alpha\beta}\Pi_{\alpha}\Pi_{\beta} + V
\end{eqnarray}
where $V = {1\over 2a}U(a) = {1\over 2a}(\lambda a^4 - a^2 + r^2_{0})$.
Here, one can readily read off the ``minisuperspace metric'' $G_{\alpha\beta}$
as 

\begin{eqnarray}
 G_{\alpha\beta} = -a\delta_{\alpha\beta}\quad,\quad
 G^{\alpha\beta} = -{1\over{a}}\delta^{\alpha\beta}
\end{eqnarray}
with $\gamma^{\alpha} = a$ and $\Pi_{\alpha} = p_{a}$ being the 
minisuperspace variable and its conjugate momentum respectively.  
Then, by the usual substitution, 

\begin{eqnarray}
 G^{\alpha\beta}\Pi_{\alpha}\Pi_{\beta}\quad\rightarrow\quad -\nabla^{2}
\end{eqnarray}
with $ \nabla^{2} = {1\over\sqrt{G}}{{\partial}\over{\partial}\gamma^{\alpha}}
(\sqrt{G}G^{\alpha\beta}{{\partial}\over{\partial}\gamma^{\beta}}) 
= -{1\over{a}}{{\partial}\over{\partial}a}({{\partial}\over{\partial}a})$, 
finally we arrive at the WD equation

\begin{eqnarray}
 {\hat H_{0}}\Psi = 
 {1\over2}\Bigl[{1\over{a}}{{\partial}\over{\partial}a}
 ({{\partial}\over{\partial}a}) + {1\over{a}}U(a)\Bigr]\Psi[a] = 0.
\end{eqnarray}
Note, here, that the minisuperspace metric (or ``DeWitt metric'')
$G_{\alpha\beta}(\gamma)$ is 
generally a function of minisuperspace variables $\gamma^{\alpha}$.  
Therefore, in passing from classical to quantum version, there arises the 
``ambiguity in operator ordering'' problem.  And as have been suggested by
Hartle and Hawking [12], one may partly encompass this ``operator-ordering''
problem by rewritting

\begin{eqnarray}
 {{\partial}\over{\partial}a}({{\partial}\over{\partial}a})
 \quad\rightarrow\quad
 {1\over{a^{p}}}{{\partial}\over{\partial}a}(a^{p}{{\partial}\over{\partial}a}).
\end{eqnarray}
Finally, the WD equation takes the general form

\begin{eqnarray}
 {1\over2}\Bigl[{1\over{a^{p}}}{{\partial}\over{\partial}a}
 (a^{p}{{\partial}\over{\partial}a}) + U(a)\Bigr]\Psi[a] = 0
\end{eqnarray}
where ``$p$'' denotes a suffix representing the ambiguity in 
operator-ordering. Before we go on and attempt to solve this WD equation explicitly, 
we would like to have some insight 
into the behavior of the solution of this WD equation. To do so, we begin by 
assigning the ``normal'' sign to the ``kinetic'' energy term to get

\begin{eqnarray}
 {1\over2}\Bigl[{-1\over{a^{p}}}{{\partial}\over{\partial}a}
 (a^{p}{{\partial}\over{\partial}a}) + \tilde{U}(a)\Bigr]\Psi[a] = 0.
\end{eqnarray}
Then the ``potential'' energy can be identified with

\begin{eqnarray}
 \tilde{U}(a) = -U(a) = (a^{2} - {\lambda}a^{4} - r^2_{0}).
\end{eqnarray}
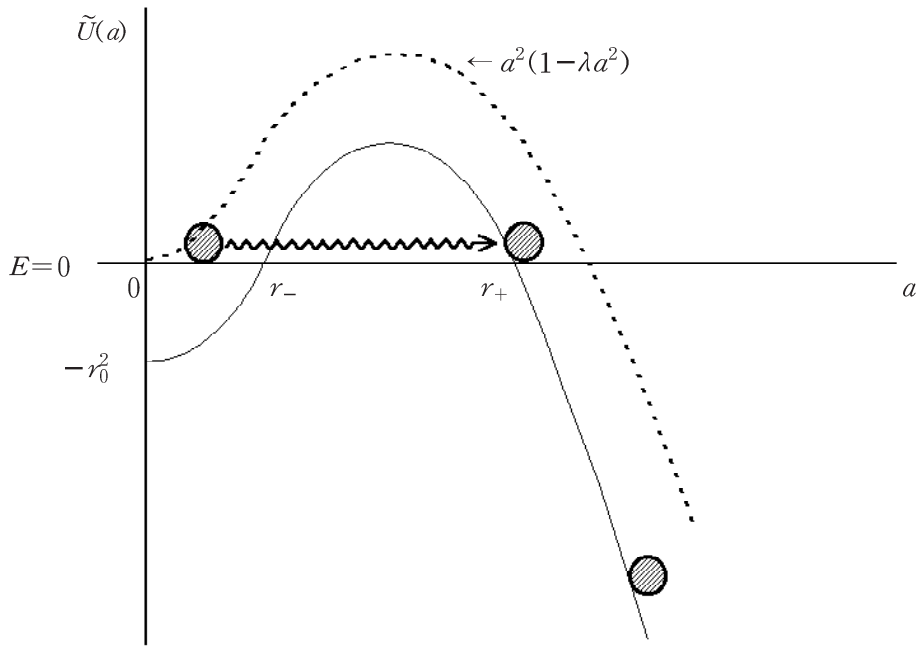
The Fig. 1(2) given displays the plot of potential energy as it appears 
in the WD equation in the presence (absence) of the cosmological constant 
$\lambda = 16\Lambda/9M^{4}_{p}$. Note here that $\tilde{U}(a) =
(a^2 - \lambda a^4 - r^2_{0}) = 0$ possesses two positive roots $R_{\pm}$
provided $\lambda < 1/4r^2_{0}$ or $\Lambda < ({g_{c}\over 8\pi})^2
{3\over 2k^2 (k+1)^2}M^4_{p}$.
Both figures show that due to the 
contribution to the potential energy, $(-r^2_{0})$, coming from 
the YM gauge field sector of the theory, the potential develops 
a ``well'' in the small-$a$ region for both cases with or without 
the cosmological constant term.  Since the WD equation implies that the 
total energy of the gravity-matter system is zero, $E = 0$, the emergence 
of the potential well in the small-$a$ region reveals the fact 
that the universe wave function $\Psi[a]$ should be an oscillating 
function of $a$ there.  And this small-$a$ behavior of the universe wave 
function appears to signal 
the existence of ``quantum wormhole'' as well as other types of spacetime 
fluctuations in the small-$a$ region of the minisuperspace and hence seems 
consistent with the existence of classical wormhole solution in this 
EYM theory as we have seen in the earlier 
section.   Obviously, the most straightforward way of confirming the 
possible existence of ``quantum wormholes'' is to solve the WD equation given 
above for the universe wave function which particularly represents an
excitation corresponding to a wormhole state.
Unfortunately, exact, analytic 
solutions to the WD equation in the presence of the cosmological constant 
are not available. (As a matter of fact, exact solutions to the WD equation 
even for de Sitter 
spacetime pure gravity are not available either [12,13].) However, since
our major concern is to find the wormhole wave function that is expected 
to exist within the potential well in the small-$a$ region, first we focus
on the case in which the cosmological constant is absent and thus the
potential energy takes the behavior depicted in Fig.2. Namely we consider
the WD equation in the absence of the cosmological constant

\begin{eqnarray}
 \Bigl[{{\partial^{2}}\over{\partial}a^{2}} 
 + {p\over{a}}{{\partial}\over{\partial}a} 
 - a^{2} + r^2_{0}\Bigr]\Psi[a] = 0.
\end{eqnarray}  
Even in this simpler case, exact, analytic solutions to this WD equation 
which is valid for the whole range of $a$, $[0,\infty )$ are not available.
However, this is not so disappointing since in order to see if this WD
equation really admits a solution that can be interpreted as  a wormhole
wave function satisfying, say, the Hawking and Page-type wormhole boundary
conditions, it suffices for us to know just the asymptotic behaviors
(such as those at small-$a$ or at large-$a$) of the solution. Thus in the
following we consider the behavior of the solution in the small-$a$ region
and in the large-$a$ region (relative to the scale $r_{0}$ set by the
dimensionless coupling parameter of the theory, i.e., 
$r^2_{0} = k^2(k+1)^2 6\pi^2 / g^2_{c}$). \\
(i) For $a << r_{0}$ : \\
In this small-$a$ region, the WD equation in the absence of the cosmological
constant above can be approximated by
\begin{eqnarray}
 \Bigl[{{\partial^{2}}\over{\partial}a^{2}}
 + {p\over{a}}{{\partial}\over{\partial}a} + r^2_{0}\Bigr]\Psi[a] \approx 0
\end{eqnarray}
which admits an analytic solution given by 
\begin{eqnarray}
\Psi[a << r_{0}] = a^{({1-p \over 2})}J_{({1-p \over 2})}(r_{0}a).
\end{eqnarray}
Note that this solution in the small-$a$ region is {\it regular} at $a = 0$
provided $p < 1$ and it represents an oscillating solution. \\
(ii) For $a >> r_{0}$ : \\
In this large-$a$ region, the WD equation above reduces to
\begin{eqnarray}
 \Bigl[{{\partial^{2}}\over{\partial}a^{2}}
 + {p\over{a}}{{\partial}\over{\partial}a} - a^2\Bigr]\Psi[a] \approx 0
\end{eqnarray}
which admits an analytic solution given by
\begin{eqnarray}
\Psi[a >> r_{0}] = a^{({1-p \over 2})}J_{({1-p \over 4})}({i\over 2}a^2).
\end{eqnarray}
Note that as $a\rightarrow \infty$, $\Psi[a >> r_{0}] \rightarrow
 a^{-({1+p \over 2})}e^{\pm {1\over 2}a^2}$, thus this solution does
involve a damping solution in the large-$a$ region. \\
Namly the solution to the WD equation in the absence of the cosmological
constant possesses regular, oscillating behavior in the classically-allowed
region, $a << r_{0}$, while it damps rapidly enough in the classically-forbidden
region, $a >> r_{0}$ in accordance with our expectation. Then next, in order 
to see if these asymptotic behaviors of the solution are consistent with or
equivalent to  the ``wormhole boundary condition'' and hence allows us to
identify the solution with the wave function of a quantum wormhole, we now
refer to a wormhole boundary condition advocated by Hawking and Page [10].
Hawking and Page argued in a convincing manner that in order for a solution to 
a WD equation to represent quantum wormholes, it should obey certain boundary
conditions. And the appropriate boundary conditions for wormhole wave
functions seem to be that they are damped, say, expnentially for large
3-geometries ($\sqrt{h} \rightarrow \infty$) and are regular in some suitable
way when the 3-geometry collapses to zero ($\sqrt{h} \rightarrow 0$). 
Particularly, in the context of FRW minisuperspace model, large 3-geometries
correspond to $a\rightarrow $large limit and the 3-geometry collapsing to 
zero corresponds to $a\rightarrow 0$ limit. And the damping behavior of the
universe wave function at large-$a$ indicates that there are no gravitational
excitations asymptotically and hence it represents asymptotically-Euclidean
spacetime while its regularity at $a = 0$ indicates that it is non-singular.
Therefore, the solution to a WD equation obeying these boundary conditions
must correspond to wormholes that connect two asymptotically-Euclidean regions.
As we already have observed, since our solution to the WD equation in the 
absence of the cosmological constant is indeed regular at $a = 0$ (provided
$p < 1$) and damps sufficiently rapidly at large-$a$, it does satisfy the 
``wornhole boundary condition''. Namely the WD equation in the absence of
the cosmological constant does admit wormhole wave function as its solution
and most probably this will hold true for the WD equation in the presence of
the cosmological constant. And this is because the essential point that 
allows the existence
of the wormhole wave function in the small-$a$ region is the contribution to
the potential energy, $(-r^2_{0})$, coming from the YM gauge field sector of 
the theory as we mentioned earlier. Therefore, now we can conclude that there
are quantum wormholes as well as classical wormhole solutions in EYM theory
with or without the cosmological constant. \\
Next we go back to the case where the cosmological constant is present. We now
would like to point out that the EYM theory with the cosmological constant
may serve as an interesting simple model which provides an overall picture
of universe's entire history from the deep quantum domain all the way to
the typical, classical domain. To this end, first note that
our knowledge on the nature of the 
universe wave function in the absence of the cosmological constant developed 
thus far may, in turn, enables us to construct the solution to the WD equation 
in the presence of the cosmological constant at least approximately yet quite 
systematically. Let us go back and consider the WD equation in 
EYM theory in the presence of the cosmological 
constant given earlier
 
\begin{eqnarray}
 {1\over2}\Bigl[-{1\over{a^{p}}}{{\partial}\over{\partial}a}
 (a^{p}{{\partial}\over{\partial}a}) 
 + (a^{2} - {\lambda}a^{4} - r^2_{0})\Bigr]\Psi[a] = 0.
\end{eqnarray}     
With the potential energy, $\tilde{U}(a) = (a^{2} - r^2_{0})$, 
as it appears in the WD equation in the absence of the cosmological constant, 
we now know that the solution to the WD equation represents 
a quantum wormhole states with oscillating behavior in the small-$a$ region.
Therefore this 
observation plus the shape of the full potential energy, $\tilde{U}(a) = 
(a^{2} -{\lambda}a^{4} - r^2_{0})$ in the presence of the 
cosmological constant as was depicted in Fig.1 suggest that the solution 
to the WD equation above would describe the state of the universe that 
undergoes ``spacetime fluctuations for very small-$a$''$\rightarrow$ 
``spontaneous nucleation (quantum tunnelling) of the universe from nothing 
in a de Sitter geometry'' $\rightarrow$ ``subsequent, mainly classical 
evolution of the universe for large-$a$''.  Thus if we are willing to 
accept (${\partial}/{\partial}a$) as the timelike killing field in the 
(mini)superspace, the EYM theory with the non-vanishing 
cosmological constant appears to serve as a simple yet interesting 
model which provides a comprehensive overall picture of entire universe's 
history from the deep quantum domain all the way to the essentially classical 
domain.  Then coming back to a practical problem, now we wish to construct 
the approximate solutions to the WD equation.  Roughly, the behavior of the 
solution for very small-$a$ will be determined by the wormhole wave function 
obtained in the present work while the behavior for intermediate-to-large-$a$ 
regions will be approximately governed by the de Sitter space universe 
wave function.  
Here, the de Sitter space universe wave functions have 
been constructed and extensively studied in the literature associated with
the issue of initial or boundary conditions for the universe wave function.
Therefore in order to employ the de Sitter space universe wave function, 
here we give a brief account of the issue of initial or boundary conditions
for the universe wave function. The WD equation is a second order hyperbolic
functional differential equation describing the evolution of the universe
wave function in the superspace. Thus the WD equation, in general, has a
large number of solutions and in order to have any predictive power, one
needs initial or boundary conditions to pick out just one solution by,
for instance, giving the value of the universe wave function at the 
boundary of the superspace. Thus far, a number of different proposals for 
the law of initial or boundary conditions have been put forward. And among
them, ``no-boundary proposal'' of Hartle and Hawking (HH) [12] and ``tunnelling
boundary condition'' due to Vilenkin [13] are the ones which are the most
comprehensive and the most extensively studied. If stated briefly, the 
no boundary proposal by HH [12] is based on the philosophy that the quantum
state of the universe is the closed cosmology's version of ``ground state''
or ``state of minimum excitation'' and the wave function of this ground
state is given by an Euclidean sum-over-histories. Next, Vilenkin's 
tunnelling boundary condition [13] can be best stated in ``outgoing modes''
formulation which governs the behavior of the solutions to the WD equation
at boundaries of the superspace. Namely, according to this proposal,
at ``singular'' boundaries (such as the region of zero 3-metric and
infinite 3-curvature ($\sqrt{h} \rightarrow 0$)) of superspace, the 
universe wave function should consist solely of outgoing modes carrying
flux out of superspace. These two proposals for the law of initial or
boundary conditions on the universe wave function determine the behaviors 
of the universe wave function like how the universe nucleated (from
``nothing'') and then following which line it has subsequently evolved
to the present one. Indeed in the minisuperspace models and within the 
context of the semiclassical approximation, these two proposals have
been successfully applied to and tested for simple systems such as
de Sitter spacetime in pure gravity which is of our interest here or
a scalar field theory coupled to gravity. \\
Thus employing the well-known de Sitter space universe wave functions
with the choice of HH's no boundary proposal or Vilenkin's tunnelling
boundary condition, we now construct the approximate solutions to
the WD equation. And to do so we consider the WD equation in two regions
of interest in the minisuperspace. \\
Firstly for very small-$a$ region, the WD equation reduces to eq.(51).
As we have investigated earlier, the important asymptotic behaviors 
of the solution to this equation are given by

\begin{eqnarray}
 \Psi_{I}[a] 
     &\rightarrow& a^{({{1-p}\over2})}J_{({1-p \over 2})}(r_{0}a)\qquad 
         ({\rm for}\quad a\rightarrow 0) \\
     &\rightarrow& a^{({{1-p}\over2})}J_{({1-p \over 4})}({i\over 2}a^2)
         \rightarrow  a^{-({{1+p}\over2})}e^{\pm {1\over 2}a^2}\qquad 
         \qquad({\rm for}\quad a\rightarrow {\rm large})\nonumber
\end{eqnarray}
where $p < 1$ for regularity of the solution.  Secondly for 
intermediate-to-large-$a$ regions, the WD equation can be roughly 
approximated by

\begin{eqnarray}
 \Bigl[{{\partial^{2}}\over{\partial}a^{2}} 
 + {p\over{a}}{{\partial}\over{\partial}a} 
 - a^{2} + {\lambda}a^{4}\Bigr]\Psi[a] = 0.\nonumber  
\end{eqnarray}
As mentioned, the semiclassical approximation to the solutions of this de 
Sitter space WD equation has been thoroughly studied.  First HH's no-boundary 
wave function [12] is given by
 
\begin{eqnarray}
 \Psi^{HH}_{II}[a]
 &=& a^{-({{1+p}\over2})}\exp{\Bigl[{1\over3\lambda}
     \{1-(1-{\lambda}a^{2})^{3/2}\}\Bigr]}\qquad
     ({\rm for}\quad{\lambda}a^{2}<1)\nonumber\\
   &&\rightarrow a^{-({{1+p}\over2})}e^{{1\over2}a^{2}}
     \quad({\lambda}a^{2}<<1), \\
 &=& a^{-({{1+p}\over2})}\exp{[{1\over3\lambda}]}
     2cos\Bigl[{({\lambda}a^{2} - 1)^{3/2}\over3\lambda} - {\pi\over4}\Bigr]
     \qquad({\rm for}\quad{\lambda}a^{2}>1)\nonumber\\
   &&\rightarrow a^{-({{1+p}\over2})}
     \Bigl[e^{i{\sqrt{\lambda}\over3}a^{3}} 
     + e^{-i{\sqrt{\lambda}\over3}a^{3}}\Bigr]
     \quad({\lambda}a^{2}>>1).\nonumber
\end{eqnarray}
This HH's no-boundary wave function consists of both ``ingoing''(contracting) 
and ``outgoing''(reexpanding) modes in the classical-allowed region 
(${\lambda}a^{2}>1$) which, then, decreases exponentially as it moves towards 
smaller values of $a$ in the classically-forbidden region (${\lambda}a^{2}<1$).
Next, Vilenkin's tunnelling wave function [13] is given by

\begin{eqnarray}
 \Psi^{T}_{II}[a]
 &=& a^{-({{1+p}\over2})}(1-{\lambda}a^{2})^{-1/4}
     \exp{\Bigl[-{1\over3\lambda}\{1-(1-{\lambda}a^{2})^{3/2}\}\Bigr]}\qquad
     ({\rm for}\quad{\lambda}a^{2}<1)\nonumber\\
   &&\rightarrow a^{-({{1+p}\over2})}e^{-{1\over2}a^{2}}
     \quad({\lambda}a^{2}<<1), \\
 &=& a^{-({{1+p}\over2})}e^{i{\pi\over4}}({\lambda}a^{2}-1)^{-1/4}
     \exp{\Bigl[-{1\over3\lambda}\{1+i({\lambda}a^{2}-1)^{3/2}\}\Bigr]}\qquad
     \qquad({\rm for}\quad{\lambda}a^{2}>1)\nonumber
\end{eqnarray}
This Vilenkin's tunnelling wave function exponentially decreases as it moves 
from small toward larger values of $a$ (i.e., emerges out of the potential 
barrier via ``quantum tunnelling'') and then upon escaping the barrier, 
it consists solely of ``outgoing'' (expanding) mode in the classically-allowed 
region (${\lambda}a^{2}>1$). Now we are ready to write down approximate 
solutions to the WD equation in the presence of the cosmological constant by 
putting these pieces altogether.  To do so let us denote the smaller and 
larger roots of the equation

\begin{eqnarray}
 \tilde{U}(a) = a^{2} - {\lambda}a^{4} - r^2_{0} = 0\nonumber
\end{eqnarray}
by $r_{-}$ and $r_{+}$ respectively which exist provided $\lambda <1/4r^2_{0}$
or equivalently $\Lambda < {3\over 2}[{g_{c}\over 8\pi k(k+1)}]^2 M^4_{p}$. 
Then the two roots are given by $r^{2}_{\pm} = {1\over 2\lambda} 
[1\pm\sqrt{1-4\lambda r^2_{0}}]$. \\        
(1) With the choice of HH's no-boundary wave function :

\begin{eqnarray}
 \Psi[a]
 &=& \Psi_{I}[a]\qquad\qquad({\rm region}\ I: 0<a<r_{+}),\nonumber\\
 &=& \Psi^{HH}_{II}[a]\qquad\qquad({\rm region}\ II: r_{-}<a<{\infty}).
 \nonumber
\end{eqnarray}
(2) With the choice of Vilenkin's tunnelling  wave function :

\begin{eqnarray}
 \Psi[a]
 &=& \Psi_{I}[a]\qquad\qquad({\rm region}\ I: 0<a<r_{+}),\nonumber\\
 &=& \Psi^{T}_{II}[a]\qquad\qquad({\rm region}\ II: r_{-}<a<{\infty}).
 \nonumber
\end{eqnarray}
The two kinds of universe wave functions corresponding to the two different choices of the boundary conditions are plotted in Fig 3. and 4 respectively.

\centerline {\rm \bf IV. Discussions}

Now we summarize the motivation and the results of the present work. \\
We revisited, in this work, the Einstein-Yang-Mills theory
considered first by Hosoya and Ogura [5] which is one of the classic systems
known to admit classical, Euclidean wormhole instanton solution. 
The classical wormhole instanton in this theory as a solution to the classical
field equations and much of its effects on low energy physics have been
studied extensively in the literature. 
Since this EYM system admits classical wormhole
solutions, one may wonder if there is any systematic way of exploring
the existence and the physics of ``quantum'' wormholes in the same
theory. The present work attempted to deal with this kind of yet
unquestioned issue.  In order to explore the
quantum wormholes in this system systematically, we worked in the
context of canonical quantum cosmology and followed Hawking and Page [10]
to define the quantum wormhole as a state or an excitation represented
by a solution to the Wheeler-DeWitt equation satisfying a certain 
wormhole boundary condition. Particularly, in the minisuperspace quantum
cosmology model possessing  SO(4)-symmetry, it is demonstrated that there
indeed exists a solution to the Wheeler-DeWitt equation whose asymptotic 
behaviors satisfy the appropriate wormhole
boundary condition for the case when the cosmological
constant is absent. Thus we confirmed our expectation that the EYM
system does admit quantum wormholes as well as classical
wormholes. Further, we pointed out that the minisuperspace quantum
cosmology model based on this EYM theory
in the presence of the cosmological constant may serve as an simple
yet interesting system displaying an overall picture of entire
universe's history from the deep quantum domain all the way to the
classical domain. \\
As we have stressed in the text, the essential point that allowed us to
explore, in a concrete manner, the quantum wormhole in the context of the
minisuperspace quantum cosmology model was the following observation.
In their original work, Hosoya and Ogura [5] obtained the expressions
for the YM gauge potential and its field strength tensor only after
explicitly solving the classical YM field equation. In particular, the
SO(4)-symmetric YM field strength tensor obtained in this way,
$F^{a}_{0b}=0$, $F^{a}_{bc}={k(k+1)\over \sigma^2 a^2(\tau)}\epsilon^{abc}$,
in turn, reduces the Einstein equation to that of the scale factor
$a(\tau)$ alone. However, we realized in this work that even without imposing
the on-shell condition (i.e., the classical field equation), one can ``derive''
$F^{a}_{bc} = {const.\over a^2}\epsilon^{abc}$ just
from the definitions and the Bianchi identity.
Therefore this SO(4)-symmetric ans\H atz for the YM
field strength $F^{a}_{bc}$ remains valid even off-shell as well as
on-shell and hence can be used in the quantum treatment of the 
EYM system. Consequently, the 
Wheeler-DeWitt equation in the context of the canonical quantum cosmology
becomes a Schr\H odinger-type equation of the minisuperspace variable
$a$ (the scale factor) alone and can in principle be solved particularly in
the absence of the cosmological constant. \\
Finally, our study of quantum wormholes in this work appears to demonstrate
that, after all, the EYM theory is a simple
yet fruitful system which serves as an arena in which we can 
envisage quite a few exciting aspects of quantum gravitational phenomena.

\centerline {\rm \bf Acknowledgements}

This work was supported in part by Korea Research Foundation and by Basic
Science Research Institute (BSRI-97-2427) at Ewha Women's Univ.

\newpage
\vspace*{2cm}

\centerline {\bf \large References}

\begin{description}

\item {[1]} E. Baum, Phys. Lett. {\bf B133}, 185 (1983) ; 
            S. W. Hawking, Phys. Lett. {\bf B134}, 403 (1984).
\item {[2]} S. Giddings and A. Strominger, Nucl. Phys. {\bf B307},
            854 (1988).
\item {[3]} S. Coleman, Nucl. Phys. {\bf B307}, 864 (1988) ;
            Nucl. Phys. {\bf B310}, 643 (1988) ; I. Klebanov, L. Susskind,
            and T. Banks, Nucl. Phys. {\bf B317}, 665 (1989). 
\item {[4]} S. Giddings and A. Strominger, Nucl. Phys. {\bf B306}, 890
            (1988) ; S. -J. Rey, Nucl. Phys. {\bf B319}, 765 (1989).
\item {[5]} A. Hosoya and W. Ogura, Phys. Lett. {\bf B22}, 117 (1989) ;
            A. K. Gupta et al., Nucl. Phys. {\bf B333}, 195 (1990) ;
            S. -J. Rey, Nucl. Phys. {\bf B336}, 146 (1990).

\item {[6]} K. Lee, Phys. Pev. Lett. {\bf 61}, 263 (1988).

\item {[7]} B. Grinstein and M. B. Wise, Phys. Lett. {\bf B212}, 407 (1988) ; 
            B. Grinstein, Nucl. Phys. {\bf B321}, 439 (1989).
\item {[8]} W. Fischler and L. Susskind, Phys. Lett. {\bf B217}, 48 (1989).

\item {[9]} G. W. Gibbons, S. W. Hawking, and M. J. Perry, Nucl. Phys. 
             {\bf B138}, 141 (1978).
\item {[10]} S. W. Hawking and D. N. Page, Phys. Rev. {\bf D42}, 2655 (1990).

\item {[11]} C. W. Misner, K. S. Thorne, and J. A. Wheeler, {\it Gravitation}
             (W. H. Freeman, San Francisco, 1973).
\item {[12]} J. B. Hartle and S. W. Hawking, Phys. Rev. {\bf D28}, 2960 
             (1983) ; S. W. Hawking, Nucl. Phys. {\bf B239}, 257 (1984).

\item {[13]} A. Vilenkin, Phys. Lett. {\bf B117}, 25 (1982) ; Phys. Rev.
             {\bf D30}, 509 (1984) ; Nucl. Phys. {\bf B252}, 141 (1985) ;
             Phys. Rev. {\bf D33}, 3560 (1986) ; Phys. Rev. {\bf D37},
             888 (1988).

\end{description}

\end{document}